\documentclass{jltp}

\title{Strong-Coupling Effects in "Dirty" Superfluid $^3He$ }

\author{ G. Baramidze  and G. Kharadze }

\address{ Andronikashvili Institute of Physics, Georgian Academy of Sciences,\\
6 Tamarashvili St., 380077, Tbilisi, Georgia \\}

\runninghead{ G. Baramidze and G. Kharadze }{Strong-Coupling Effects in
"Dirty" Superfluid $^3He$ }

\begin{document}

\maketitle

\begin{abstract}

The contribution of the strong-coupling effects to the free energy
of the "dirty" superfluid $^3He$ is estimated using a simple model.
It is shown that the strong-coupling effects are less susceptible to    
the quasiparticle scattering events in comparison to the weak-coupling
counterpart. This supports the conclusion about stabilization
of the $B$-phase in aerogel environment at pressures where the $A$-phase
takes over in bulk superfluid $^3He$, in accordance with
recent experimental observations in zero magnetic field.
     
\end{abstract}     

\section{Introduction}
     
One of the actively investigated problems in the low temperature physics is
the search of the properties of a ''dirty'' superfluid Fermi system like
liquid $^{3}He$, confined to a high-porosity aerogel environment. A number
of experimental observations [1-10] revealed the new aspects of the behavior
of the ordered state of the superfluid $^{3}He$ in the presence of the
quasiparticle scattering against a random system of silica strands forming
the skeleton of aerogel.

The theoretical attempts to interpret these experiments, although partly
successful, are still insufficient to describe the main body of accumulated
information and the details of the phase diagram, in particular. The
theoretical approach adopted up to now is based on a weak-coupling
approximation. One of the conclusions obtained in this way is the claim
that in a zero magnetic field the quasiparticle scattering on the
spatial irregularities ("impurities") promotes the stability of the
isotropic $B$-phase in the domain of the $P-T$ phase diagram where in
bulk ("pure") superfluid $^3He$ the anisotropic $A$-phase takes over [11,12].
This theoretical result means that at pressures above the polycritical
value $P_{c0}$ the "dirty" $B$-phase overcomes the so called strong-coupling
effects and should appear as an equilibrium superfluid state of liquid
$^3He$ confined to the aerogel environment. This conclusion is based on a
supposition that the strong-coupling effects (which also are subject to
the "impurity" renormalization) are less susceptible to the quasiparticle
scattering events.

In what follows we explore this question in some details. It will be shown
(using a simple model) that indeed it seems likely that, although the
strong-coupling effects are enhanced due to the finite value of the
quasiparticle mean free path, the $B$-phase still is able to take over
the $A$-phase at high pressures. Quite recently, using a high frequrncy
acoustic technique [13] the 3D phase diagram in ($P,T,B$) space was constructed
for superfluid $^3He$ confined to 98$\%$ porosity aerogel (see, also, Ref.[14]).
The covered pressure range extended from 15 bars up to the melting pressure.
One of the most striking observations is that in zero magnetic field ($B=0$)
and at all pressures above 15 bars the phase transition to the $B$-like phase
takes part (at $T_{c}(P)$ with no signs of polycritical
point (PCP) at wich the $A$- and $B$-phases do meet in bulk superfluid $^3He$
at $P_{c0}$=21 bars)). It appears that PCP for $^3He$ in 98{\%} porosity aerogel
is absent because $P_c$ is pushed out above the solidification pressure, and
thus is unobservable. The results of our theoretical consideration seem to
be in accordance with the above-mentioned experimental observations.

\section{Strong-Coupling Effects in "Dirty" Superfluid $^3He$}

The weak-coupling approach in treating the properties of superfluid
phases of liquid $^3He$ disregards the inverse action of the
ordering on the 
pairing iteraction (so called strong-coupling effects). The importance of
this feedback effect is well known [15] and is mainly due to an attractive
contribution of the spin exitations in the strongly correlated system to the
effective quasiparticle interaction. Physically the strong-coupling
feedback effect stems from the fact that the spin susceptibility of liquid $%
^{3}He$ is sensitive to the character of the spin-triplet Cooper pairing
order parameter. In what follows, using a simple model, we are going to
estimate the influence of the quasiparticle scattering on the
strong-coupling effects having in mind an application to the superfluid $%
^{3}He$ filling the low-density aerogel. It should be stressed that a
much more rafined aproach in treating the strong-coupling effects is based
on a systematic expansion of the free energy of superfluid $^3He$ in powers
of $T_{c}/T_{F}$. This program has been realized in Ref.[16] (see, also,
the review article [17] and Ref.[18]). This approach captures the contributions
to the strong-coupling effects stemming not only from the spin fluctuations
but from the transverse current fluctuations as well. Unfortunately, it is
an extremely difficult task  to treat the "impurity" effects even within relatively
simple dynamical spin-fluctuation model used in Ref.[19], nothing to say about a more
sophisticated approach mentioned above. Instead, we will rely on a static
approximation [20] which disregardes the retardation effects in treating
an attractive interaction between quasiparticales via the exchange of the
"paramagnons".

We start from the momentum space Fourier component of the static
spin-susceptibility tensor

\begin{eqnarray}
\chi _{\mu \nu }(\vec{q})=-\frac{1}{2}T 
\sum _{\omega}\sum_{k}Tr\biggl [ \stackrel{\wedge }{G}_{\omega }(%
\vec{k})\stackrel{\wedge }{\sigma }_{\mu }\stackrel{%
\wedge }{G}_{\omega }(\vec{k}+\vec{q})%
\stackrel{\wedge }{\sigma }_{\nu }-\\ \nonumber
\stackrel{\wedge }{\stackrel{\_}{F}}%
_{\omega }(\vec{k})\stackrel{\wedge }{\sigma }_{\mu }%
\stackrel{\wedge }{F}_{\omega }(\vec{k}+
\vec{q})\stackrel{\wedge }{\sigma }_{\nu }^{tr}\biggr] ,
\end{eqnarray}
where $2\times 2$ spin-matrices $\stackrel{\wedge }{G}_{\omega }(\vec{k})$ and
$\stackrel{\wedge }{F}_{\omega }(\vec{k})$ denote the Gorkov Green's functions (in the Matsubara
representation), describing an ordered (superfluid) Fermi system. Eq.(1) can
be used to construct an effective spin-depended part of the interaction
potential acting between quasiparticles with the matrix elements

\begin{equation}
J_{\mu \nu }(\vec{k},\vec{k^{\prime }%
})(\stackrel{\wedge }{\sigma }_{\mu })_{\alpha \alpha ^{\prime }}(\stackrel{%
\wedge }{\sigma }_{\nu })_{\beta \beta ^{\prime }},
\end{equation}
where

\begin{equation}
J_{\mu \nu }(\vec {k},\vec {k^{\prime }%
})=-I^{2}\chi _{\mu \nu }(\vec {k}-\vec{%
k^{\prime }})
\end{equation}
with $I$ standing for the local repulsive potential describing correlation
effects.

In the random-phase approximation the susceptibility tensor $\stackrel{%
\wedge }{\chi }$ is constracted as

\begin{equation}
\stackrel{\wedge }{\chi }=\stackrel{\wedge }{\chi }^{(0)}+\stackrel{\wedge }{%
\chi }^{(0)}I\stackrel{\wedge }{\chi }=(\stackrel{\wedge }{1}-I\stackrel{%
\wedge }{\chi }^{(0)})^{-1}\stackrel{\wedge }{\chi }^{(0)},
\end{equation}
where $\stackrel{\wedge }{\chi }^{(0)}$stands for the spin susceptibility in
the absence of the correlation effects $(I=0)$. In the vicinity of the critical
temperature of the phase transition to the superfluid state

\begin{equation}
\stackrel{\wedge }{\chi }^{(0)}\simeq {\chi }_{N}^{(0)}%
\stackrel{\wedge }{1}+\delta \stackrel{\wedge }{\chi }^{(0)}
\end{equation}
with ${\chi }_{N}^{(0)}$being the normal-state
susceptibility. The superfluid contribution $\delta \stackrel{\wedge }{\chi }%
^{(0)}$is quadratic in the order parameter of the superfluid state.
Consequently, at $T\stackrel{<}{\sim}T_{c}$

\begin{equation}
J_{\mu \nu }\simeq -\frac{I^{2}{\chi }_{N}^{(0)}}{1-I%
{\chi }_{N}^{(0)}}\delta _{\mu \nu }-\left( \frac{I}{1-I%
{\chi }_{N}^{(0)}}\right) ^{2}\delta \stackrel{\wedge }{%
\chi }^{(0)}_{\mu\nu}.
\end{equation}

The results concerning the strong-coupling contribution in $^3He$
and based on the mentioned simple model are described in Ref.[21].
They show that this model reflects the essence of the feedback
effects at least qualitatively. At the same time, in the framework
of the adopted static model the technical side of the calculation
of the quasiparticle scattering effects (which is our main goal)
is simplified considerably. Hopefully the static model, which
we adopted, gives at least qualitatively meaningful treatment
of the {\it relative} stability of the $A$- and $B$-phases of a "dirty"
superfluid $^3He$ confined to aerogel environment in zero
magnetic field.

In order to estimate the effects of the finite mean free path of the
quasiparticles we address a self-consistancy equation for the
order parameter matrix $\stackrel{\wedge }{\Delta}$ which in the case of a
spin-triplet Cooper pairing (appropriate to the superfluid $^3 He$)
is given as

\begin{equation}
\stackrel{\wedge}{\Delta}=\Delta_{\mu}(\stackrel{\wedge}{\sigma}_{\mu}i
\stackrel{\wedge}{\sigma}_y).
\end{equation}
The equation for the vector component $\Delta_{\mu}$ reads as

\begin{equation}
\Delta_{\mu}(\hat{k})=-T\sum_{\omega}\sum_{k'}V_{\mu\nu}
(\vec{k},\vec{k'})F_{\omega\nu}(\vec{k'}),
\end{equation}
where $\hat{k}$ is the unit vector along the momentum $\vec{k}$,

\begin{equation}
V_{\mu\nu}=V\delta_{\mu\nu}+\delta V_{\mu\nu},
\end{equation}
and the feedback contribution $\delta V_{\mu\nu}=J_{\lambda\lambda}
\delta_{\mu\nu}-2J_{\mu\nu}.$ In eq.(8) $F_{\omega\nu}$ denotes the
$\nu$-th vector component connected to $\hat{F}_{\omega}$ in a
way similar to Eq.(7). In the adopted model

\begin{equation}
\delta V_{\mu\nu}=-\left(\frac{I}{1-I\chi_{N}^{(0)}}\right)^2
(\delta\chi_{\lambda\lambda}^{(0)}\delta_{\mu\nu}-
2\delta\chi_{\mu\nu}^{(0)}).
\end{equation}

In terms of the quasiclassical function

\begin{equation}
f_{\omega\nu}(\hat{k})=\frac{1}{\pi}\int_{-\propto}^{+\propto}
d\xi F_{\omega\nu}(\hat{k},\xi)
\end{equation}
the self-consistency Eq.(8) reads as

\begin{equation}
\Delta_{\mu}(\hat{k})=2\pi T\sum_{\omega>0}\sum_{k'}\langle
3\hat{k}\hat{k'}g_{\mu\nu}f_{\omega\nu}(\hat{k'})\rangle,
\end{equation}
where the brackets denote the averaging over the orientation of $\hat{k'}$
and $g_{\mu\nu}$ stands for an attractive component in the p-wave channel.
According to our consideration $g_{\mu\nu}=g\delta_{\mu\nu}+\delta g_{\mu\nu}$
with $\delta g_{\mu\nu}$ stemming from the spin-dependent part of the
quasiparticle interaction (see Eq.(10)).

For the case of bulk superfluid $^3 He$ near $T_{c0}$ (the critical temperature
of a pure system) and at $q\ll k_{F}$

\begin{equation}
\delta\chi^{(0)}_{\mu\nu}(\vec{q})=-2\pi T\sum_{\omega>0}
\frac{N_{F}}{2\omega^3}\left(\frac{2\omega}{qv_{F}}\right)^2
\Biggl\langle\frac{\rm{Re}(\Delta_{\mu}(\hat{k})\Delta_{\nu}^{*}(\hat{k}))}
{(\hat{k}\hat{q})^{2}+(2\omega/qv_{F})^2}\Biggr\rangle,
\end{equation}
where $N_{F}$ and $v_{F}$ denote the density of states and
the velocity of the  quasiparticles at the Fermi level. After averaging over the orientation of $\hat{q}$, from Eq.(13) it is obtained that

\begin{equation}
\delta\chi^{(0)}_{\mu\nu}(q)=-\frac{N_{F}}{qv_{F}}\left[
2\pi T~\sum_{\omega>0}
\frac{1}{\omega^2}\rm{arctan}(qv_{F}/2\omega)\right]\rm{Re}\langle
\Delta_{\mu}(\hat{k})\Delta_{\nu}^{*}(\hat{k})\rangle.
\end{equation}

The main contribution to the feedback coupling constant $\delta g_{\mu\nu}$
is to be extracted from the region of $q\gg \xi_{c0}^{-1}$ 
where the coherence length $\xi_{c0}=v_F/2\pi T_{c0}$.
In this limit Eq.(14) gives

\begin{equation}
\delta\chi^{(0)}_{\mu\nu}(q)=-\frac{\pi}{2}\frac{N_F}{qv_F}
\left(2\pi T\sum_{\omega>0}\frac{1}{\omega^2}\right)\rm{Re}
\langle\Delta_{\mu}(\hat{k})\Delta_{\nu}^{*}(\hat{k})\rangle
\end{equation}
and as a result

\begin{equation}
\delta g_{\mu\nu}=\frac{\delta g}{(\pi T_{c0})^2}
\langle |\vec{\Delta}|^2\delta_{\mu\nu}-\Delta_{\mu}\Delta_{\nu}^{*}-
\Delta_{\mu}^{*}\Delta_{\nu}\rangle,
\end{equation}
with

\begin{equation}
\delta g=\frac{1}{6}\left(\frac{\pi}{2}\right)^3\left(
\frac{IN_F}{1-IN_F}\right)^2\frac{1}{k_F\xi_{c0}}.
\end{equation}

Our main concern is to establish the modification of Eq.(16) caused by the
quasiparticle scattering against the irregularities ("impurities")
introduced by the presence of aerogel silica stands. The spin
susceptibility is a two-particle correlator and the corresponding
system of equations is to be addressed. The
impurity scattering effects show up, in particular, as the vertex
corrections complicating considerably the general consideration.
Fortunately at $\xi^{-1}_{c}\ll q \ll k_F$, which is the region of
the momentum transfer we are interested in, the vertex corrections
are small as far as $(k_{F}\xi_{c})^{-1}\ll 1$. Finally, the
essential contribution to the feedback effect modification due to
the scattering events are simply realized by the substitution
$\omega\rightarrow\tilde{\omega}=\omega+\Gamma sgn~\omega$ in
Eq.(15), where $\Gamma=\frac{c}{\pi N_F}\rm{sin}^2\delta_0$
is the quasiparticles scattering rate
(in what follow we adopt the so called homogeneus scattering
model (HSM) with the $s$-wave scattering
channel only (see Refs.[11,12]). Here $c$ denotes the "impurity"
concentration and $\delta_0$ is the phase shift at an $s$-wave
scattering. As a result the coupling constant
$\delta g_{\mu\nu}$ (see Eq.(16)) is transformed to

\begin{equation}
\delta\tilde{g}_{\mu\nu}=\frac{\delta\tilde{g}}{(\pi T_c)^2}
\langle |\vec\Delta|^2\delta_{\mu\nu}-\Delta_{\mu}\Delta_{\nu}^{*}-
\Delta_{\mu}^{*}\Delta_{\nu}\rangle,
\end{equation}
where $T_c$ stands for the critical temperature of the phase
transition of liquid $^3 He$ in aerogel to an ordered state and

\begin{equation}
\delta\tilde{g}=\frac{1}{6}\left(\frac{\pi}{2}\right)^3
\left(\frac{IN_F}{1-IN_F}\right)^2\frac{1}{k_F\xi_c}
\frac{\psi^{(1)}(1/2+w)}{\pi^2/2},~~\xi_c=\frac{v_F}{2\pi T_c}.
\end{equation}
Here $\psi^{(m)}(z)$ denotes the poly-gamma function and
$w=\Gamma/2\pi T$.

Taking into account that up to the third order in $\vec{\Delta}$
and in the presence of the quasiparticle scattering centers

\begin{eqnarray}
\vec{f}_{\omega} \simeq \frac{\vec\Delta}{|\tilde{\omega}|}-
\frac{1}{2|\tilde{\omega}|^3}\biggl [ 
(\vec{\Delta}^{*}\vec{\Delta})\vec{\Delta}+
(\vec{\Delta}^{*}\times\vec{\Delta})\times\vec{\Delta}-\\ \nonumber
\frac{\Gamma\rm{cos}2\delta_0}{|\tilde{\omega}|}\left(\langle
|\vec{\Delta}|^2\rangle\vec{\Delta}+\langle
(\vec{\Delta}^{*}\times\vec{\Delta})\rangle\times\vec{\Delta}\right)
\biggr], 
\end{eqnarray}
after averaging over the orientation of
$\hat{k}'$ in the self-consistency Eq.(12), the equation
for the order parameter $\vec{\Delta}(\hat{k})$ near $T_c$
reads as (for the superfluid 
$^3He~~\Delta_{\mu}(\hat{k})=A_{\mu i}\hat{k}_{i}$):

\begin{eqnarray}
(a_{1}(T)-1/g)\vec{\Delta}(\hat{k})=\frac{3}{5}a_3 \biggl(
-\frac{1}{2}\langle\vec{\Delta}^2\rangle \vec{\Delta}^{*}(\hat{k})+
\langle|\vec{\Delta}|^2\rangle\vec{\Delta}(\hat{k})+\\ \nonumber
\langle\vec{\Delta}\Delta_{\nu}\rangle\Delta_{\nu}^{*}(\hat{k})+
\langle\vec{\Delta}\Delta_{\nu}^{*}\rangle\Delta_{\nu}(\hat{k})-
\langle\vec{\Delta}^{*}\Delta_{\nu}\rangle\Delta_{\nu}(\hat{k})\biggr)-\\ \nonumber
\frac{1}{2}\Gamma\rm{cos}2\delta_0
a_4\left(\langle|\vec{\Delta}|^2\rangle\vec{\Delta}(\hat{k})+
\langle\vec{\Delta}\Delta_{\nu}^{*}\rangle\Delta_{\nu}(\hat{k})-
\langle\vec{\Delta}^{*}\Delta_{\nu}\rangle\Delta_{\nu}(\hat{k}\right)+\\ \nonumber
\frac{\delta\tilde{g}}{g}\frac{a_1}{(\pi T_c)^2}
\left(\langle|\vec{\Delta}|^2\rangle\vec{\Delta}(\hat{k})-
\langle\vec{\Delta}\Delta_{\nu}^{*}\rangle\Delta_{\nu}(\hat{k})-
\langle\vec{\Delta}^{*}\Delta_{\nu}\rangle\Delta_{\nu}(\hat{k})\right),
\end{eqnarray}
where

\begin{eqnarray}
a_{1}(T)=2\pi T\sum_{\omega>0}^{\omega_c}\frac{1}{\tilde{\omega}}=
ln\left(\frac{2\gamma}{\pi}\frac{\omega_c}{T}\right)
+\psi(1/2)-\psi(1/2+w),\\ 
a_{3}(T)=2\pi T\sum_{\omega>0}\frac{1}{\tilde{\omega}^3}=
-\frac{1}{2}\frac{1}{(2\pi T)^2}\psi^{(2)}(1/2+w),\\ 
a_{4}(T)=2\pi T\sum_{\omega>0}\frac{1}{\tilde{\omega}^4}=
\frac{1}{6}\frac{1}{(2\pi T)^3}\psi^{(3)}(1/2+w).
\end{eqnarray}

Finally, for the order parameter $A_{\mu i}$ the following equation is
obtained from Eq.(21)

\begin{eqnarray}
\alpha (T)A_{\mu i}+\frac{N_F}{15}a_3\biggr\{ 
-\frac{1}{2}A_{\mu i}^{*}A_{\nu j}A_{\nu j}+\\ \nonumber
\left[ \left(
1-\frac{5}{6}\Gamma\rm{cos}2\delta_0\frac{a_4}{a_3}\right)+\delta_{sc}\right]
A_{\mu i}A_{\nu j}^{*}A_{\nu j}+\\ \nonumber
A_{\mu j}A_{\nu j}A_{\nu i}^{*}+
\left[ \left(
1-\frac{5}{6}\Gamma\rm{cos}2\delta_0\frac{a_4}{a_3}\right)-\delta_{sc}\right]
A_{\mu j}A_{\nu j}^{*}A_{\nu i}-\\ \nonumber
\left[ \left(
1-\frac{5}{6}\Gamma\rm{cos}2\delta_0\frac{a_4}{a_3}\right)+\delta_{sc}\right]
A_{\mu j}^{*}A_{\nu j}A_{\nu i}\biggr\}=0,\nonumber
\end{eqnarray}
where $\alpha(T)=\frac{1}{3}N_F\left[\rm{ln}\frac{T}{T_{c0}}+\psi(1/2+\omega)-
\psi(1/2)\right]$ and the strong-coupling contribution is described
by a parameter

\begin{equation}
\delta_{sc}=\frac{5}{3}\frac{\delta \tilde{g}}{g^2}\frac{1}{a_3}
\frac{1}{(\pi T_c)^2}.
\end{equation}
In Eqs.(25) and (26) the coefficients $a_3$ and $a_4$ are to be calculated
at $T=T_c$.

Comparing Eq.(25) with its phenomenological Ginzburg-Landau counterpart

\begin{eqnarray}
\alpha (T)A_{\mu i}+2(\beta_{1}A_{\mu i}^{*}A_{\nu j}A_{\nu j}+
\beta_{2}A_{\mu i}A_{\nu j}^{*}A_{\nu j}+\\ \nonumber
\beta_{3}A_{\mu j}A_{\nu j}A_{\nu i}^{*}+
\beta_{4}A_{\mu j}A_{\nu j}^{*}A_{\nu i}+
\beta_{5}A_{\mu j}^{*}A_{\nu j}A_{\nu i})=0,
\end{eqnarray}
the $\beta$-coefficients can be identified. Representing $\beta_i$ as the
sum of week-coupling ($\beta_i^{wc}$) and strong-coupling
($\delta\beta_i^{sc}$) contributions, it is found that

\begin{eqnarray}
-2\beta_1^{wc} =\beta_3^{wc}=2\beta_{wc}=
\frac{7\zeta(3)}{120}\frac{N_F}{(\pi T_c)^2}
\frac{\psi^{(2)}(1/2+w_c)}{\psi^{(2)}(1/2)},\\
\beta_2^{wc}=\beta_4^{wc}=-\beta_5^{wc}=
2\beta_{wc}-\frac{\Gamma\rm{cos}2\delta_0}{12^3}
\frac{N_F}{(\pi T_c)^3}\psi^{(3)}(1/2+w_c),\\
\delta\beta_1^{sc}=\delta\beta_3^{sc}=0,\\
\delta\beta_2^{sc}=-\delta\beta_4^{sc}=-\delta\beta_5^{sc}=
\delta\beta_{sc},\\
\delta\beta_{sc}=\frac{1}{2(4g)^2}\left(\frac{\pi}{3}\right)^3
\frac{N_F}{(\pi T_c)^2}
\left(\frac{IN_F}{1-IN_F}\right)^2\frac{1}{k_F\xi_c}
\frac{\psi^{(1)}(1/2+w_c)}{\psi^{(1)}(1/2)},
\end{eqnarray}
where $w_c=\frac{\Gamma}{2\pi T_c}$. It can be easily verified that the
weak-coupling coefficients $\beta_i^{wc}$ reproduce the answer reported in 
Ref.[11].

In order to explore the domain of the phase diagram (in the
Ginzburg-Landau region) where the $B$-phase overcomes the strong-coupling
effects and is preferable as an equilibrium state in comparision to
the $A$-phase (in zero magnetic field), we address a well known inequality

\begin{equation}
\beta_{12}+\frac{1}{3}\beta_{345}<\beta_{245}.
\end{equation}

Introducing the normalized $\beta$-coefficients
$\bar{\beta}_i=\beta_i/|\beta_1^{wc}|$, the criterion of thermodynamical
stability of the $B$-phase in the Ginzburg-Landau region reads as

\begin{equation}
-2\delta\bar{\beta}^{sc}_{345}+3\delta\bar{\beta}^{sc}_{13}<1.
\end{equation}

According to Eq.[28] $|\beta^{wc}_1|=\beta _{wc}^{0}R_{wc}$ where the
"impurity" renormalization factor for the weak-coupling coefficient
$\beta _{wc}$ is given by

\begin{equation}
R_{wc}(w_c)=\frac{\psi^{(2)}(1/2+w_c)}{\psi^{(2)}(1/2)}
\left(\frac{T_{c0}}{T_c}\right)^2.
\end{equation}

On the other hand, following Eq.(32)
$\delta\beta_{sc}=\delta\beta_{sc}^{0}R_{sc}$ with the "impurity"
renormalization factor

\begin{equation}
R_{sc}(w_c)=\frac{\psi^{(1)}(1/2+w_c)}{\psi^{(1)}(1/2)}
\frac{T_{c0}}{T_c}.
\end{equation}

It is to be remembered that the ratio $T_{c0}/T_{c}(w_c)$ is found
from the Abrikosov-Gorkov type equation

\begin{equation}
\rm{ln}(T_{c0}/T_c)+\psi(1/2)-\psi(1/2+\omega_c)=0.
\end{equation}

The renormalization factors $R_{wc}$ and $R_{sc}$ are monotonously
increasing functions of $w_c$ although the strong-coupling effects
are less susceptible to the quasiparticle scattering events.

Collecting these results, in the framework of the adopted simple
model the $B$-phase stability region near $T_c$ is defined by the inequality

\begin{equation}
\delta\bar{\beta}_{sc}=\delta\bar{\beta}_{sc}^{0}(P)R(w_{sc})<\frac{1}{4},
\end{equation}
where the effects of the finite mean free path of the quasiparticles in
aerogel environment are accumulated in the renormalization factor

\begin{equation}
R(w_c)=\frac{R_{sc}(w_c)}{R_{wc}(w_c)}=a(w_c)\frac{T_c}{T_{c0}}
\end{equation}
with

\begin{equation}
a(w_c)=\frac{\psi^{(1)}(1/2+w_c)}{\psi^{(1)}(1/2)}
\frac{\psi^{(2)}(1/2)}{\psi^{(2)}(1/2+w_c)}.
\end{equation}

It is to be noted that in Eq.(38) the presence of $T_c/T_{c0}$ certainly
stems from the fact that the strong-coupling corrections to the free energy
contain an extra powers in $T_c/T_F$ (in comparison with the weak-coupling
contribution). At the same time, Eq.(38) shows that $R(w_c)$ is not simply equal
to $T_c/T_{c0}$ but contains an extra factor $a(w_c)$ which increases with the quasiparticle scattering rate and competes with $T_c/T_{c0}$ which decreases with $w_c$. The analyses shows that this
competition is in favor of $T_c/T_{c0}$ and $R_{wc}<1$ at $w_c>0$. In particular, for the case with $w_c<<1~a(w_c)\simeq 1+2.37w_c$ and
$T_c/T_{c0}\simeq 1-\frac{1}{2}\pi^2 w_c$, so that $R(w_c)\simeq 1-2.56w_c$.

Turning back to Eq.(37) it is concluded that, since in the quasiparticle
scattering medium $R(w_c)<1$, the condition of the stability of the $B$-phase
in aerogel is less restrictive in comparison with bulk case. This opens
a way for the appearence of the $B$-like superfluid state in the pressure
region $P>P_{c0}$ which increases with the quasiparticle scattering intensity.
As was mentioned in the Introduction, in case of 98$\%$ porosity aerogel the
$B$-like phase near $T_c$ and in zero magnetic field is observed up to the
melting pressure $P_m$. For larger porosity aerogel (with smaller $w_c$)
the PCP may appear in the pressure region $P_{c0}<P<P_m$, as mentioned in
Ref.[13].

\section{Conclusion}

As is well known, an isotropic $B$-phase of superfluid $^3He$ is stabilized
at pressures below $P_{c0}\simeq 21$ bars. At higher pressures the $A$-phase takes over
due to the strong-coupling effects manifested as a feedback of the Cooper pairing
on the quasiparticle attractive interaction.

On the other hand, in recent experimental studies (see Ref.[13]) of
the phase diagram of a "dirty" superfluid $^3He$ confined to
the 98$\%$ porosity silica aerogel,
it was established that a $B$-phase-like ordered state is stabilized at
$P>P_{c0}$ up to $P=P_m$. This observation indicates that the scattering of
quasiparticles against the spatial irregularities of the porous medium
modifies the free energy of superfluid $^3He$ in favor of the $B$-phase
as an equilibrium ordered state at high pressures.
The free energy of superfluid state can be viewed as containing the two
contributions stemming from the weak-coupling and strong-coupling
effects. The former contribution for the "dirty" superfluid $^3He$
has been investigated theoretically in Ref.[11] where it was shown
that the "impurity" renormalization of the week-coupling part of the
free energy (near $T_c$) promotes the stabilization of the $B$-phase
at the pressures where in bulk superfluid $^3He$ the A-phase is an
equilibrium ordered state. This conclusion, as mentioned in Ref.[11],
supposes that the strong-coupling effects are not more susceptible to
the quasiparticle scattering then their weak-coupling counterpart.
In using a simple model to treat the strong-coupling contribution to
the free energy, we have demonstrated that the "impurity"
renormalization of this contribution is considerably weaker in
comparison to the weak-coupling effects. This conclusion seems
to be in accordance with the mentioned experimental observations.

\section*{ACKNOWLEDGMENTS}

We are grateful to Prof. Bill Halperin for informing us about the
results described in Ref.[13].
This work was supported by the INTAS Grant  N 97-1643 and the Georgian
Academy of Sciences Grant N 2-19.

\end{document}